\documentclass{pnastwo}

\usepackage{bm}

\url{www.pnas.org/cgi/doi/10.1073/pnas.0709640104}
\copyrightyear{2008}
\issuedate{Issue Date}
\volume{Volume}
\issuenumber{Issue Number}

\begin{document}

\title{Topological patterns in two-dimensional gel electrophoresis of DNA knots}

\author{Davide Michieletto\affil{1}{Department of Physics and Complexity Science, University of Warwick, Coventry,
United Kingdom},
Davide Marenduzzo\affil{2}{School of Physics and Astronomy, University of Edinburgh, Edinburgh, United Kingdom},
\and
Enzo Orlandini\affil{3}{Dipartimento di Fisica e Astronomia and Sezione INFN, Universit\`a di Padova, Padova, Italy.}}




\contributor{Submitted to Proceedings of the National Academy of Sciences
of the United States of America}

\significancetext{Gel electrophoresis is a ubiquitous biophysical technique. It consists in dragging charged biopolymers through a porous gel, by applying an electric field. Because the migration speed depends on topology, this method can be used to classify DNA knots.
Currently, electrophoresis largely relies on empirical observations, while its theoretical understanding is limited. For instance, no theory can explain why knot mobility under strong fields depends non-monotonically on complexity. Our study reveals a possible reason: complex knots have a smaller size, hence move more quickly through the gel pores; however, once they get trapped by the dangling ends present in the gel, the resulting entanglements are slower to resolve. Our results can improve the design of future electrophoresis experiments.
}

\maketitle

\begin{article}
\begin{abstract}
Gel electrophoresis is a powerful experimental method to probe the topology of DNA and other biopolymers. While there is a large body of experimental work which allows us to accurately separate different topoisomers of a molecule, a full theoretical understanding of these experiments has not yet been achieved. Here we show that the mobility of DNA knots depends crucially and subtly on the physical properties of the gel, and in particular on the presence of dangling ends. The topological interactions between these and DNA molecules can be described in terms of an ``entanglement number'', and yield a non-monotonic mobility at moderate fields. Consequently, in two-dimensional electrophoresis, gel bands display a characteristic arc pattern; this turns into a straight line when the density of dangling ends vanishes. We also provide a novel framework to accurately predict the shape of such arcs as a function of molecule length and topological complexity, which may be used to inform future experiments.
\end{abstract}

\keywords{DNA Knots | Gel Electrophoresis | Topology }

\dropcap{T}opology plays a key role in the biophysics of DNA, and is intimately related to its functioning. For instance, transcription of a gene redistributes twist locally to create what is known as supercoiling, while catenanes or knots can prevent cell division, hence they need to be quickly and accurately removed by specialised enzymes known as topoisomerases. But how can one establish experimentally the topological state of a given DNA molecule? By far the most successful and widely used technique to do so is gel electrophoresis~\cite{Calladine1997,Bates2005}. This method exploits the empirical observation that the mobility of a charged DNA molecule under an electric field depends on its size, shape and topology~\cite{Bates2005}. Gel electrophoresis is so reliable that it can be used, for instance, to map replication origins and stalled replication forks~\cite{Olavarrieta2002}, to separate plasmids with different amount of supercoiling~\cite{Olavarrieta2002,Cebrian2014}, and to identify DNA knots~\cite{Stas,Arsuaga2005}. The most widely employed variant of this technique nowadays is two-dimensional gel electrophoresis, where a DNA molecule is subjected to a sequence of two fields, applied along orthogonal directions~\cite{Bates2005}. The two runs are characterised by different field strengths, and sometimes also gel concentrations~\cite{Cebrian2014}; with suitable choices, the joint responses leads to increased sensitivity.

While gel electrophoresis is used very often, and is extremely well characterised empirically, there is still no comprehensive theory to quantitatively understand, or predict, what results will be observed in a particular experiment. 
Some aspects are reasonably well established. For instance, it is now widely accepted that the physics of the size-dependent migration of linear polymers can be explained by the theory of biased polymer reptation~\cite{Gennes1979,Rubinstein1987,Duke1989,Viovy1993,Barkema1994,Viovy2000}. Likewise, the behaviour of, for example, nicked, torsionally relaxed, DNA knots in a sparse gel and under a weak field is analogous to that of molecules sedimenting under gravity~\cite{Weber2013a,Piili2013}. The terminal velocity can be estimated via a balance between the applied force and the frictional opposing force, which is proportional to the average size of the molecule: as a result more complex knots, which are smaller, move faster under the field.
However, the mechanisms regulating the electrophoretic mobility of DNA knots at intermediate fields, and in more concentrated agar gels, is much less understood~\cite{Cebrian2014,Viovy2000,Weber2006a}. Here, experiments suggest that the mobility of knots is usually a non-monotonic function of the knot complexity, or, more precisely, of their average crossing number~\cite{Katritch1996a,Stas} (ACN): initially knots move more slowly as their ACN increases, while, past a critical ACN, more complex knots  move faster. The combination of the responses to external fields directed along two perpendicular directions leads to a characteristic electrophoretic arc which allows to separate the first simple knots more clearly in a 2D slab~\cite{Trigueros2001,Arsuaga2002,Arsuaga2005,Cebrian2014}. To our knowledge, there is currently no theoretical framework that quantitatively explains the non-monotonic behaviour at intermediate or large fields and the consequent formation of arc patterns. 

\begin{figure}[h]
\centering
\includegraphics[width=0.35\textwidth]{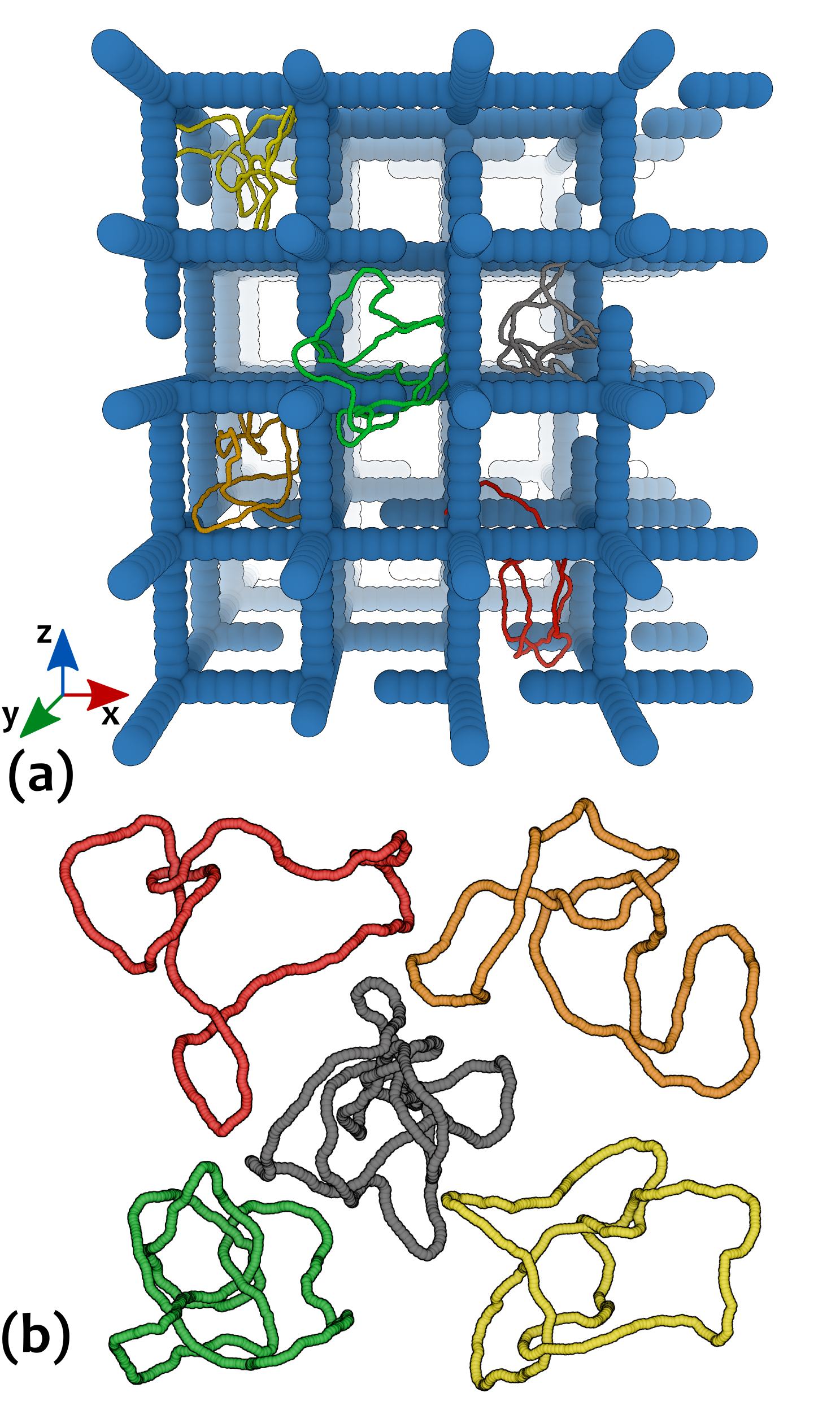}
\caption{\textbf{(a)} Snapshot (to scale) of the model gel with some examples of knotted configurations. Note that to model a physical  gel, a simple cubic structure is randomly cut to create dangling ends. \textbf{(b)} Equilibrium configurations of some of the knots considered; it can be readily seen that the size tends to be smaller as the knot becomes more complex. The knots pictured in \textbf{(a)} and \textbf{(b)} are: trefoil ($3_1$) in red, figure of eight ($4_1$) in orange, pentafoil ($5_1$) in yellow, Stevedore's ($6_1$) in green and ``nonafoil'' ($9_1$) in grey.}
\label{fig:Panel1}
\end{figure}

To address this issue, here we present large scale Brownian dynamics simulations of knotted DNA chains migrating through a gel, and subjected to a sequence of fields of different strength and direction, as in two-dimensional gel electrophoresis experiments (see cartoon in Fig.~\ref{fig:Panel1}(a)). We model the gel as an imperfect cubic mesh~\cite{Michieletto2014d}, where some of the bonds have been cut (see Methods) to simulate the presence of open strands, or dangling ends, which have been observed in physical agarose gels~\cite{Turmel1990,Cole2002,Cole2003a,Robertson2007,Stellwagen2009}. 
Our results confirm the linear relation of the electrophoretic mobility with average crossing number for the first simple knots (we study ACN up to 12) in a sparse gel and under a weak field. 
However, our simulations also suggest that, due to a non-negligible probability of forming ``impalements'' where a dangling end of the gel pierces a knot, the response of the chain to stronger fields is different. We suggest that, in this regime, the sole radius of gyration is not enough to explain the observed dynamics, and we introduce an average ``entanglement number'' which increases with the ACN and provides a measure of the likelihood of forming an impalement.  The time needed by a knot to disentangle from the gel increases with its average entanglement number, or knot complexity, and this slows down the motion, thus competing with the Stokes friction which leads to an increase of mobility with ACN.
As a result of this contest, one typically gets a non-monotonic behaviour of the terminal speed with ACN, and an electrophoretic arc in 2 dimensions.
 

We also provide a simple model, based on a mapping between the DNA knot dynamics and a biased continuous time random walks, which faithfully reproduces our Brownian dynamics simulations starting from a minimal number of assumption. This approach can then be used to predict how the shape of the electrophoretic arc should depend on system parameters such as the average lattice spacing (pore size) of the gel and the contour length of DNA knots; as we shall see, the predicted trends are in agreement with existing two-dimensional electrophoresis data. This constitutes, to our knowledge, the first example of a quantitative prediction of two-dimensional electrophoretic diagrams, hence we suggest that the approach we present could potentially lead to even more accurate and targeted experiments to separate topoisomers in DNA or other polymers.

\subsection{DNA knots form an electrophoretic arc only in irregular gels}

The system we studied, sketched in Fig.~\ref{fig:Panel1} (see also Materials and Methods and SI), consists of 10 nicked, \emph{i.e.} torsionally relaxed, DNA loops of $\sim 3.7$ kilo-base pairs (kbp) within a model agarose gel with dangling ends. These loops are either unknotted, or form one of the first few simple knots (with up to 9 crossing in their minimal projection~\cite{Adams1994}). 
The loops were first equilibrated within the gel (see SI), and then subjected to an {\it in silico} gel electrophoresis process where a weak electric field is first applied ($\lesssim 50$ $V/cm$) along the vertical ($z$) direction, followed by a stronger field ($\gtrsim 150$ $V/cm$) along a transversal, say $y$, direction. We refer to these two fields as ``weak'' and ``stronger'', or ``moderate'', in what follows. The complete equations of motions and force fields used in our Brownian dynamics simulations are detailed in the SI.

\begin{figure*}[t]
\centering
\includegraphics[width=0.95\textwidth]{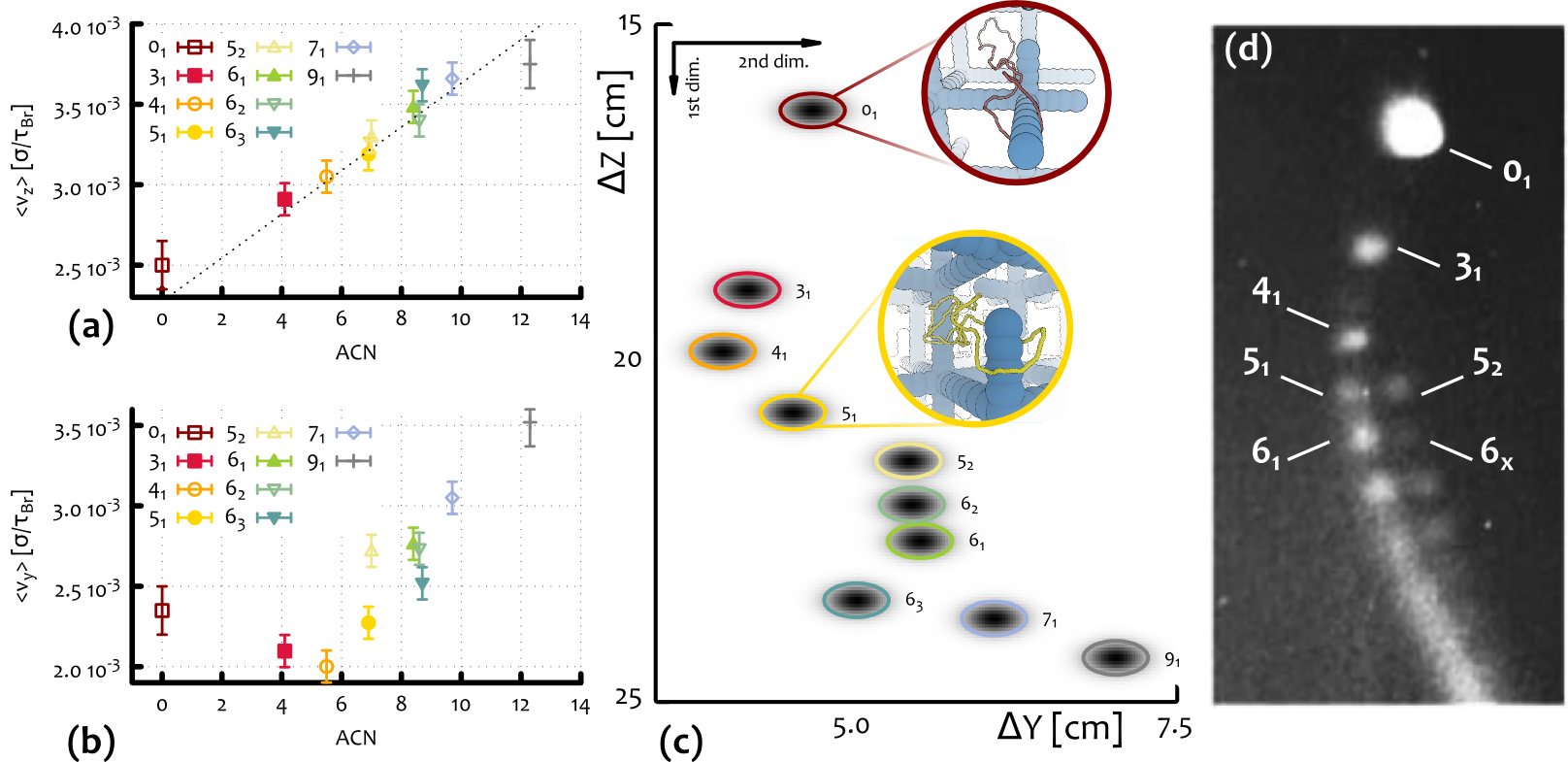}
\caption{\emph{In silico} 2D gel electrophoresis for knotted DNA loops. \textbf{(a)} Average velocity of different types of knotted polymers along the direction of the weak electric field: $E_1 = f_1/q_b \simeq  50$ $V/cm$. The dotted line indicates the linear increase with the ACN. \textbf{(b)}  Average velocity of different types of knotted polymers along the direction of the moderate field $E_2 = f_2/q_b = 150$ $V/cm$. 
\textbf{(c)} 2D reconstruction of the spatial distribution of the knots as Gaussian probability distributions centred in ($v_z(\mathcal{K}) t_z$,$v_y(\mathcal{K})t_y$) where $v_z(\mathcal{K})$ and $v_y(\mathcal{K})$ are the average speeds along, respectively, the weak and stronger field direction of knot type $K$, as found by Brownian dynamics simulations. The electrophoretic run-times correspond to $t_z=0.8$ $10^{10}$ $\tau_{Br} \simeq 5$ $min$ and $t_y=2.6$ $10^{10}$ $\tau_{Br} \simeq 16$ $min$.
The spread of each spot has been estimated by looking at the standard deviation of $v_z(\mathcal{K})$ and $v_y(\mathcal{K})$. (Examples of single trajectories along the $z$ and $y$ directions are reported in  Fig.~S1). \textbf{(d)} Outcome of a 2D gel electrophoresis experiment performed on P4 viral DNA ($10$ $kbp$) at 0.4\% agarose concentration (reproduced from Ref.~\cite{Arsuaga2005} with permission). }
\label{fig:2DGE}
\end{figure*}

By monitoring the trajectories of the knots through the gel we computed the average speed of their center of mass along each of the field directions (see SI, and Fig.\ref{fig:2DGE}a-b). As expected the mobility along the direction of the weak field increases with the topological complexity of the configurations.  Along the direction of the moderate field, however, the mobility of the knots displays a non-monotonic behaviour. In particular, the unknot now moves faster than either the trefoil or the $4_1$ twist knot, and has an average speed similar to the $5_1$ knot. This non monotonic behaviour of the knot mobility, as a function of the ACN, was previously  observed in typical experiments with torsionally relaxed DNA knots~\cite{Trigueros2001,Arsuaga2002,Arsuaga2005,Trigueros2007,Cebrian2014}.  
It is worth noticing that, within the $5$-crossings family the $5_1$ torus knot moves more slowly than the $5_2$ twist knots. This is similar to what observed in the weak field case (although much less enhanced) but different from what is observed in experiments of sedimentation~\cite{Piili2013}.

To better compare our findings with experiments we report, in Fig.~\ref{fig:2DGE}c the spatial distribution of the knots as Gaussians centred in ($v_z(\mathcal{K}) t_z$,$v_y(\mathcal{K})t_y$) where $v_z(\mathcal{K})$ and $v_y(\mathcal{K})$ are the 
velocities along $z$ (weak field direction) and $y$ (stronger field direction) of knot $K$, while $t_y$ and $t_z$ the electrophoretic run-times. The width of the Gaussians is set to be proportional to the standard deviation of the velocities. 
The resulting spots can be seen as the {\it in silico} analogue of the ones observed in gel electrophoretic experiments. 
Note that the combination of a monotonic behaviour along the weak field direction with a non-monotonic one along the stronger field, gives rise to the arc shape distribution of the spots characteristic of 2D electrophoresis experiments run either on knotted configurations or on or supercoiled plasmids~\cite{Cebrian2014} (see Fig.~\ref{fig:2DGE}d). 

It is interesting to ask whether one can observe the electrophoretic arc also in simulations where the gel is a regular cubic mesh, \emph{i.e.} a mesh with no dangling ends, as this has been so far the typical way to model an agarose gel~\cite{Weber2006a}. Remarkably, unlike the case of gel with dangling ends, also called ``irregular'' hereafter, no example of non-monotonic behaviour of the knot mobility is found for regular gels (for comparison see Fig.~\ref{fig:2DGE}c) and Fig.~S1 in SI). This result is in line with previous simulations based on lattice knots in regular gels~\cite{Weber2006a} and persists for different field strengths ($1.25-600$ $V/cm$) and gel pore sizes ($200-500$ $nm$) (see SI).~\footnote{We note that physical gels also have an inhomogeneous pore size; while considering this aspect will affect our results quantitatively, the common understanding is that the knot speed should depend monotonically on size~\cite{Calladine1997}. This is qualitatively different from the case of dangling ends, where the gel-polymer interactions strongly depend on topological complexity as well.} Hence our simulations strongly suggests that the causes for the non-monotonic behaviour, observed in the case of irregular gels,  are to be sought in the interaction between the knots and the gel dangling ends. 

This conjecture is also supported by the fact that linear (open) DNA samples, are frequently observed to migrate faster than covalently closed (unknotted) ones, in gel electrophoresis experiments performed in both strong and weak fields~\cite{Trigueros2001,Arsuaga2005,Trigueros2007,Cebrian2014}. This is in line with the outcome of a recent computer experiment probing the dynamics of linear and unknotted circular molecules through an irregular gel~\cite{Michieletto2014d}.

\subsection{DNA samples become severely entangled with the dangling ends}

Having established that the presence of dangling ends in gels affect severely the transport properties of the knotted DNA loops under moderate electric fields, it is natural to look at the possible mechanisms ruling this phenomenon. 

The  typical trajectories and average extension of some knotted loops, as they move through a regular model gel and a 
gel with dangling ends are markedly different at moderate fields (see S.I. Fig.~S1). 
In a regular gel, knots respond to the field, by shrinking their size so as to channel through the pores of the 
gel more efficiently. This mechanism, also known as ``channelling'', for which polymers squeeze through the gel pores, 
has already been observed in previous works~\cite{Mohan2007,Mohan2007b}, and it was previously conjectured to play a role in the non-monotonic separation of DNA knots in gels, as more complex knots could have a different ability in deforming their overall shape when squeezing through the pores~\cite{Cebrian2014}. On the other hand, as discussed in the preceding section, we find that this behaviour is not sufficient to explain the electrophoretic arc, as for regular gels we always observe a monotonic separation of the knots as a function of the ACN (see S.I., Fig.~S1).
On the contrary, in the case of irregular gels, knotted loops are much more prone to entangle 
with one (or more) dangling ends (see insets of Fig.~\ref{fig:2DGE}c for some examples). 
These entangled states (or ``impalements") require some time to be unraveled and this is 
the reason of the anomalously long pauses observed in the knot trajectories (see in particular SI Fig.~S1).
Clearly, as the DNA gets longer, ``impalements'', which can either be parallel or perpendicular with respect the direction of the field, become progressively more likely. As a matter of fact
this could be one of the reasons why it is in practice unfeasible to perform efficient gel electrophoresis experiments with circular DNA longer than $10$ $kbp$~\cite{Dorfman2010}: at these sizes impalements are so frequent that they may cause DNA breakage.

In analogy with the phenomenon of threading, which slows down the dynamics of unknotted loops either in a melt 
or in a gel~\cite{Michieletto2014,Michieletto2014a,Michieletto2014d}, and that of ``crawling'' of knots around 
obstacles~\cite{Weber2006b},  it is reasonable to expect that more complicated knots 
will take longer to disentangle themselves from an impalement. 
We argue that this mechanism, when competing with the reduced Stokes drag of more complex 
knots in gels, is ultimately responsible for producing a non-monotonic dependence as a function of their 
complexity, \emph{i.e.} their ACN.

\begin{figure*}
\centering
 \includegraphics[width=0.8\textwidth]{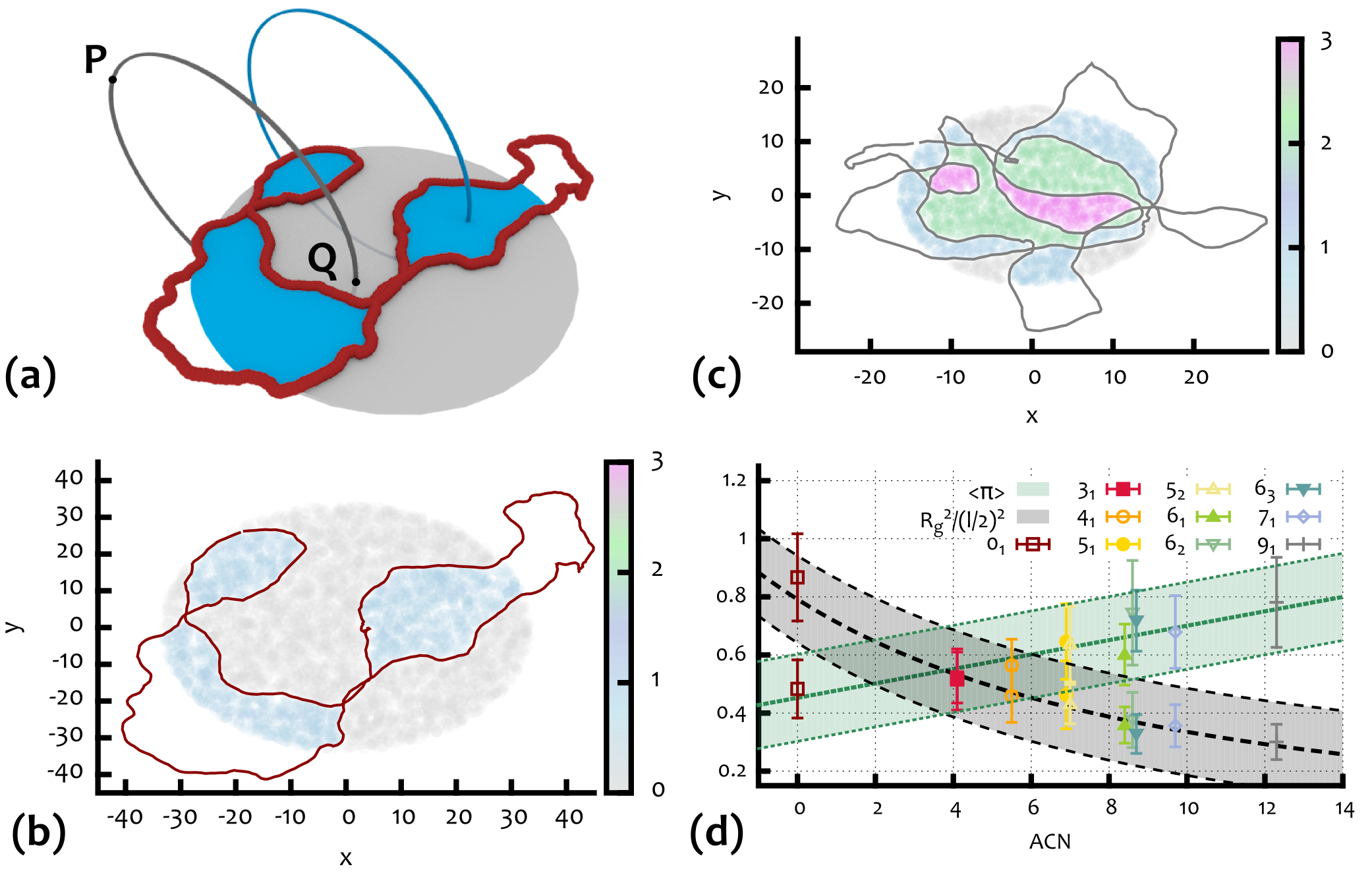}
\caption{\textbf{(a)} Sketch of the procedure used to define the piercing, or entanglement number: for a given projection of the configuration, the crossings define a set of regions whose intersections with the disk of radius $R_g^{^{proj}}$ are highlighted in blue. Starting from different points $P$ far away from the projections closed paths (2 in the example) that pierce once the disk at different locations $Q$  are built at random. The absolute value of the linking number is then computed between the knot configuration and each closed path. The average over the set of closed paths is finally taken: this is the average entanglement number.
\textbf{(b)} and \textbf{(c)} show the results of the procedure described in panel \textbf{(a)} for a configuration with knot type $9_1$ \textbf{(c)} and for an unknotted configuration \textbf{(b)}. The regions are coloured according to the computed absolute value of the linking number (see colour map at the left). Note that for the $9_1$ case there are regions of high (3) $|Lk|$, which are more prone to become entangled with the dangling ends of the gel.
\textbf{(d)} shows the average entanglement number $\langle \pi \rangle$ and the mean squared radius of gyration divided by $(l/2)^2$ ($l$ is the gel pore size) for different knot types classified in terms of ACN.}
\label{fig:Rg_and_Pi}
\end{figure*}

\subsection{More complex knots have smaller size but larger ``entanglement number''}
Given that ``impalement'' events are key factors in determining the mobility of DNA knots within gels with dangling ends, it is important to find a way to define and measure this entanglement. Impalement may occur with dangling ends oriented along several directions (see examples in Fig.~\ref{fig:2DGE}(c)) but it is reasonable to expect that all these events involve a 
similar mechanism in which, \emph{i.e.} one dangling end ``pierces through'' the knot. 

To quantify the degree of knot-gel entanglement, we consider an equilibrated knot configuration 
in the gel, and project it on the plane perpendicular to the field direction. 
We then choose randomly  a base point $P$, at a distance from the projection plane that is much bigger than the 
radius of gyration of the projected configuration, $R_g^{proj}$. 
Starting from $P$ we draw an arc which pierces only once the projection plane at a point, $Q$, chosen randomly, 
with uniform probability within a disk of radius $R_g^{proj}$ and centred in the center of mass of the projected configuration. The arc and the plane define a semispace and we  close the path with a second arc connecting $Q$ and $P$ and living 
in the other semispace (see Fig.~\ref{fig:Rg_and_Pi}a). To assess whether this circular path interacts topologically with the knotted configuration, we compute the absolute value of the linking number, $|Lk|$, between the circular path and the knot. 
Fig.~\ref{fig:Rg_and_Pi}b and Fig.~\ref{fig:Rg_and_Pi}c show the result of this procedure when applied to two 
different knotted loops.
By averaging $|Lk|$ over several circular paths with different $Q$, and over different knot configurations, 
we define the ``average entanglement number'',  $\langle \pi \rangle$ (AEN) as
the measure of the degree of entanglement between the knotted loops and the surrounding gel.


From Fig.~\ref{fig:Rg_and_Pi}d we see that $\langle \pi \rangle$  grows approximately linearly with ACN: as one would expect, 
more complex knots can, on average, become more entangled with the surrounding irregular gel. It is interesting to compare this behaviour with that of the mean squared radius of gyration normalised with respect to half the gel pore size (see Fig.~\ref{fig:Rg_and_Pi}d): unlike the average entanglement number, the average extension of the loop is, to a good approximation, 
inversely proportional to the ACN, \emph{i.e.} to the knot complexity. 
This corresponds to the well-known fact that, for a given loop contour length, more complex knots are on average less extended~\cite{Stas,Piili2013} (see also the equilibrium configurations in Fig.~\ref{fig:Panel1}). 

The plots in Fig.~\ref{fig:Rg_and_Pi}d suggest a possible interpretation of the non monotonic mobility of the knots in irregular gels based on the interplay between the average size and the degree of entanglement with the gel. On one hand more complex knots, being smaller in size, experience less frequent collision with the gel and hence should travel more easily through it: 
this is just another variant of the Stokes friction argument discussed previously. 
On the other hand,  once knot-gel collisions occur, more complex knots experience a more intricate entanglement with the 
gel (higher values of AEN are more probable) that will take longer to be unravelled~\cite{Stas,Weber2013a,Weber2006a,Weber2006b}.

 
The above argument suggests the existence of two time scales in the process:
one is the time $\tau_f$ between two successive knot-gel collisions yielding a local entanglement; 
the other, $\tau_{\rm dis}$, is the time needed by the knotted loop to fully disentangle from the impalement. 
The time scale $\tau_f$ increases as the knot average size decreases and hence increases with knot complexity 
(ACN). In other words more complex knots experience, on average, less collisions with the gel than their simpler counterpart.
The second timescale, $\tau_{\rm dis}$, is instead an increasing function of  $\langle \pi \rangle$ (see Fig.~\ref{fig:Rg_and_Pi}(d)) and hence of the knot complexity (measured in terms of ACN). 
According to this picture, the slowest topoisomer in an irregular  gel  with a given lattice spacing will be the one
with the ``best'' compromise between a high rate of collisions, and a sufficiently high value $\langle \pi \rangle$. 


\begin{figure}
 \centering
\includegraphics[width=0.45\textwidth]{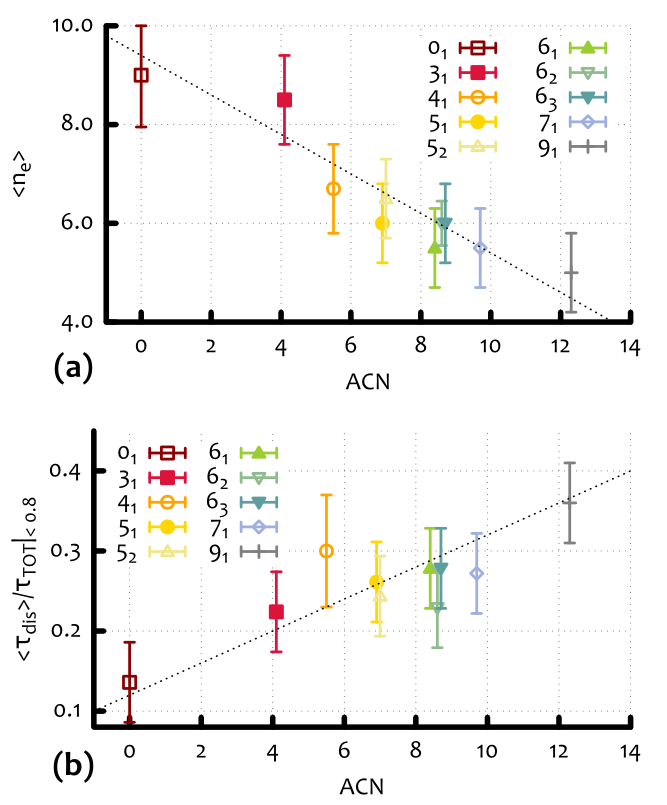}
\caption{
\textbf{(a)} Average number of events in which the knot is entangled with the surrounding gel (entanglement events) as a function of ACN. 
\textbf{(b)} Average disentanglement time as a function of ACN. In these estimates only entanglement events with duration shorter than 200 $\tau_{Br}$ are considered.}
\label{fig:Wtimes}
\end{figure}

To investigate more quantitatively the dependence of $\tau_f$ and $\tau_{\rm dis}$ on the knot type (ACN), 
we analyse the trajectories of the knotted loops in the gel by computing: (i) the average number of times a knot arrests its motion in the gel (entanglement event), $\langle n_e \rangle$, and (ii) the distribution of the duration of these entanglement events. 
As specified in the SI, the duration of the entanglement events can be identified as the time intervals 
where the spatial position of the centre of mass of the configuration deviates significantly from the 
expected collision-free field-driven linear motion with speed $v_{free} = F_{z,y}/M\zeta=f_{z,y}/\zeta$.


As reported in the SI (Fig.~S2), the average fraction of time in which the knot is trapped, $\tau_{w}/\tau_{tot}$ ($\tau_{tot}$ is the time of the full trajectory), is a non-monotonic function of the ACN, in line with the result on 
the mobility under moderate field (Fig.~\ref{fig:2DGE}b). 
In Fig.~\ref{fig:Wtimes}(a) we  show that the average number of entanglement events $\langle n_e \rangle$ decreases 
with the knot complexity (i.e. ACN). On the other hand the distribution of the duration of  these events 
displays an intriguing bimodal shape, with two peaks occurring respectively at short and very long times (see SI Fig.~S3). 
The peak at long times can be interpreted as the signature of head-on collisions with the gel, 
where the dangling ends involved  are opposite to the direction of the knot motion. 
The resulting entanglement is, in this case, very difficult to unravel especially in presence of a strong electric 
field (see for instance the inset of Fig.~\ref{fig:2DGE} for the unknot). 
Nonetheless, either  if we exclude or not  from the statistic the  entanglement events corresponding to the peak at long times
the characteristic  disentanglement time $\tau_{\rm dis}$ turns out to  increase (linearly) with the knot complexity, \emph{i.e.} with $ACN$ (see Fig.~4(b)). 

We assume that this bi-modal shape is due to a shift in the energy barrier that the knots have
to overcome in order to disentangle from the dangling ends. In particular, one can think to this process 
as an Arrhenius process, where the energy barrier is a function of the length (projected along the field direction) 
of the dangling end, the knot complexity and, more importantly, the magnitude of the external field. 
When this is too strong, disentanglement events are very rare, and all knots will end up being permanently entangled with the gel structure; on the other hand, when this is too weak, the typical disentanglement time is very short, and the dependence of $\tau_f$ as a function of the ACN dominates the motion of the polymers, re-establishing the usual linear relationship.


\subsection{A random walk model with topology-dependent rates captures the observed non-monotonic behaviour}

As shown in previous sections, 2D electrophoresis experiments and Brownian dynamics simulations of knotted loops in 
irregular gels are in qualitative agreement under many aspects. In this section we propose a 
simple model that reproduces the main findings of the simulations and furnishes a simple but accurate way 
to predict the arc shapes of the experimental patterns as a function both of the knot complexity and of the  
loop  contour length.  

In this model, we describe the knotted loop moving within the irregular gel 
as a biased random walk on a 1D lattice i.e. a random walk that 
moves to the right (direction of the external field) unless it is trapped into an 
entangled state (due to impalement) with probability $\lambda_e(\mathcal{K})=\tau_f^{-1}(\mathcal{K})$ 
(see SI for more details). 
Once in the entangled state, the walker has to wait a given amount of time that is 
picked randomly from a bimodal distribution consisting of an exponential decay, modelling the short time 
disentanglement, and a smaller probability peak at large times, describing the long disentanglement time from a head-on 
collision (see SI for the details). In this simple description the only relevant 
parameters are the hitting rate and the parameters characterising the bimodal distribution of waiting (i.e. disentanglement)
times. 

\begin{figure*}[t]
\centering
\includegraphics[width=0.9\textwidth]{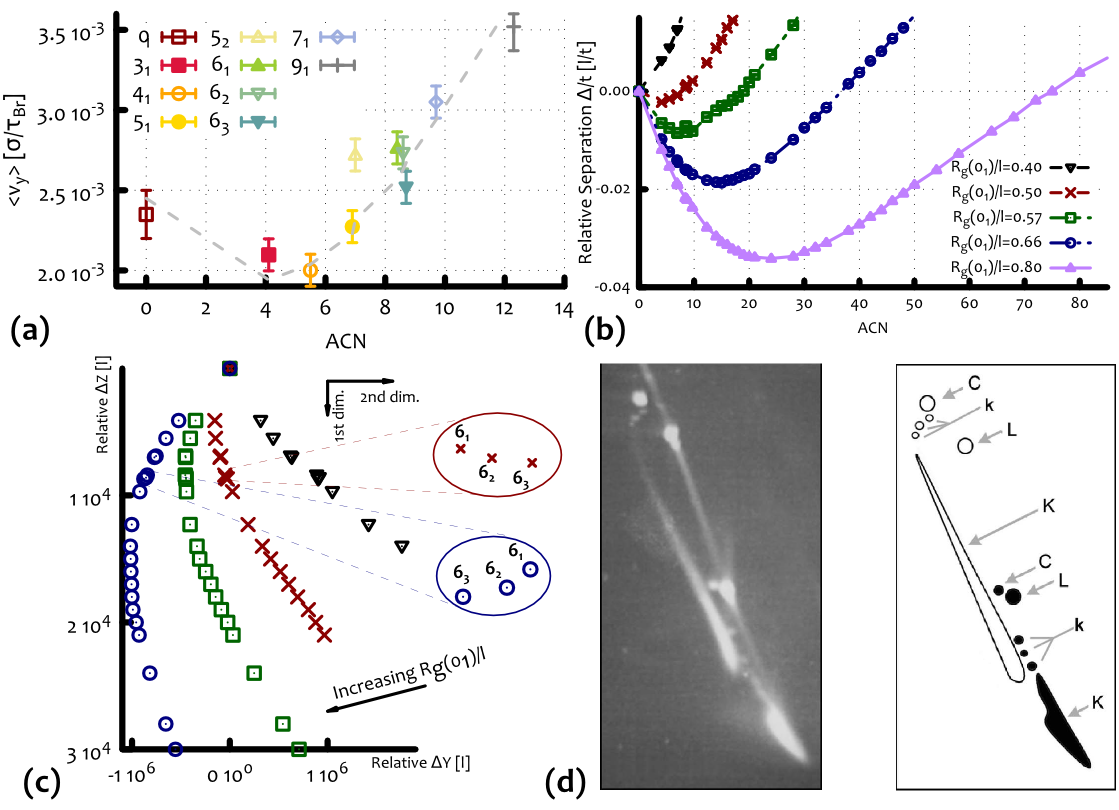}
\caption{\textbf{(a)} Average speed along the direction of the moderate field from Fig.~\ref{fig:2DGE}(b). The dashed line is obtained from the biased continuous random walk model, and corresponds to the (shifted and rescaled) red curve in \textbf{(b)}. \textbf{(b)} Average relative separation (in units of lattice spacing over time) of the knots as a function of the ACN for different parameters, as predicted by the continuous random walk model. The grey dashed line in \textbf{(a)} is obtained by shifting the red curve in \textbf{(b)} by the value of $\langle v_y \rangle(0_1)$ and rescaling it by the free velocity $v_{free}$. \textbf{(c)} Reconstruction of a 2D gel electrophoresis experiment from the data in \textbf{(b)} and zoom over the relative position of the family of 6-crossings knots for two cases in which the minimum of the arc is at their left and their right. \textbf{(d)} Outcomes of a 2D gel electrophoresis experiments performed on P4 viral DNA with different lengths, respectively $4.7$ (black) and $10$ (white) $kbp$ at equal agarose concentration (0.4 \%) (reproduced from Ref.~\cite{Trigueros2007} with permission).}
\label{fig:RW_vs_MD}
\end{figure*}

Once the values of these parameters are set to reproduce the data reported in Fig.~\ref{fig:Wtimes} (see also SI),  
the model can be used to predict the mobility of the electrophoretic arc as a function of ACN. As shown in 
Fig.~\ref{fig:RW_vs_MD} this procedure reproduces with remarkably good agreement the simulation data and, in particular,
it captures the physical mechanism leading to the non-monotonic mobility at moderate field. Note that, as the 
random walker solely moves to the right, the field strength enters into the model only through 
the waiting times and the hitting rates.


More importantly, once the parameter values of the biased random walk model are set for a given pore size of the gel, $l_1$, 
their values for a different pore size, $l_2$, can be estimated from general arguments (see SI). 
We can therefore use this simplified model to predict the moderate field mobility and the shape of the electrophoretic 
arcs of DNA knots in gels of variable pore size, \emph{e.g.} tuned via agarose concentration~\cite{Pernodet1997} or nano-wire growth cycle~\cite{Rahong2014}.

The plots presented in Fig.~\ref{fig:RW_vs_MD}(b)  suggest that tighter 
gels give rise to more curved (or deeper) arcs where the slowest knot has a higher ACN with respect to sparser gels. 
Moreover, since the entanglement rate $\lambda_e$ and the disentanglement time $\tau_{\rm dis}$ (both of these relative to the same quantities for the unknot) should  depend only on the ratio between the knot extension and the gel pore size $l$, a similar trend should be observed also by increasing the DNA loop contour length by keeping fixed $l$ (see SI).
This is in qualitative agreement with experiments, as electrophoretic arcs are straighter for shorter DNA molecules (Fig.~\ref{fig:RW_vs_MD}(d)). A further quantitative prediction we can draw from our arguments is that the relative position of the three 6-crossing knots can be controlled by tuning the pore size (Fig.~\ref{fig:RW_vs_MD}) of the gel. 
Indeed the size of the pores determines whether the $6_1$ Stevedore's knot is to the left or to the right of 
the minimum of the mobility curve: in the former case $6_1$ will moves faster in the gel 
than the $6_2$ and $6_3$ knots (which, having higher ACN have also higher AEN), while in the latter case it will
move more slowly. 
This detailed prediction could be tested in future electrophoresis experiments with knotted DNA loops moving
within different gels. 

\subsection{Conclusions}

We have studied the role of topology in the gel electrophoretic mobility of DNA knots by means of Brownian dynamics simulations and a minimal model of biased random walk. 
We showed that, when the knots are driven through a physical gel \emph{i.e.} possessing dangling ends, 
the  knots' mobility, as a function of their average crossing number (ACN), depends on the strength of the external field. 

At weak fields we recover the well-known linear relationship between migrating speed and knot-type; at stronger fields, we observe instead  a non-monotonic behaviour. We argue that this puzzling feature, routinely observed in experiments but yet not fully explained, can be better understood by taking into account the topological interactions, or entanglements, of the knots with the irregularities of the surrounding gel. While more complex knots assume more compact configurations, 
and hence smaller  Stokes friction than simpler knots, they also experience  more complex entanglements with the gel and hence longer disentanglement times. 
These two competing effects give rise to  the non-monotonic speed of the knots observed in the experiments, a feature that, remarkably, is absent for knotted loops moving in  a regular gel (\emph{i.e.} no dangling ends). While most of our simulations were performed with a rigid gel, we
tested that the results are qualitatively unchanged for gels with flexible dangling ends (see Materials and Methods, and SI).

We also propose a model which describe the motion of knotted  DNA loops as a biased continuous time random walks.
This model, although minimal, by focusing on  the competition between Stokes friction and topological entanglements 
highlighted by the simulations, is able to predict the shape of electrophoretic patterns of DNA knots of different 
contour length observed in gels with tunable physical properties. 
In particular we predict that, by changing the ratio between the radius of gyration of the unknot and the gel pore size,
2D gel electrophoresis experiments should lead to deep electrophoretic arcs for tight gels (or long knots), 
and shallow ones for sparse gels (or short knots).  

We hope that our results will prompt further experimental and numerical verification on the role of topology 
in the anomalous electrophoretic mobility of knotted polymers and, consequently, suggest new and more accurate setups 
to separate biopolymers of different topology.
Lastly, it is  likely that a similar competition between loop size and loop-gel interactions 
can be responsible of the observed  electrophoretic arcs in DNA molecules with different density of 
supercoiling; we plan to explore this related situation, as well as the case of composite knots, in the future.

\begin{materials}
Double-stranded (ds) and nicked, \emph{i.e.} torsionally relaxed, DNA knots are modelled as closed and knotted semi-flexible bead-spring chains~\cite{Kremer1990}, with beads of diameter {\small$\sigma=2.5$ nm}, which reflects the thickness of hydrated B-DNA near physiological conditions~\cite{Rybenkov1993}.  The persistence length is set to {\small$l_p = 20 \sigma= 50$ nm}, and the chosen contour length {\small$L_c=512 \sigma$} corresponds to DNA loops of length {\small$\sim 4$ $kbp$ $\simeq 1.3$ $\mu$ m}. 

The gel is modelled as an imperfect and rigid cubic mesh, with lattice spacing  {\small$l=80 \sigma \simeq 200$ $nm$} compatible with the average pore size of agarose gels at 5\% and artificial gels made of solid nano-wires~\cite{Pernodet1997,Rahong2014} (for more details on the model and comparison with the case of flexible dangling ends, see SI). The irregularities, or dangling ends, of the gel are created by starting from a regular cubic mesh and then by halving some of the edges randomly, with probability {\small$p=0.4$}. Although this probability is chosen arbitrarily, it is possible to map it to a real value of ``disorder'' found in an agarose gel at a given concentration by comparing the mobility of linear and ring polymers running through it, similarly to what was done in Ref.~\cite{Michieletto2014d}.
The edges of the mesh are discretised with beads of size {\small$\sigma_g = 10 \sigma \simeq 25$} nm which is compatible with the observed diameter of agarose bundles~\cite{Pernodet1997,Guenet2006}. For simplicity and computational efficiency, we model the gel as a static mesh, meaning that the mesh structure is not deforming under either thermal of mechanical strains. This is an approximation for an agarose gel, whose bundles are generally found, at the concentrations used in gel electrophoresis, to be made of tens of fibers whose persistence length has been observed to be around {\small$2-10$} nm~\cite{Guenet2006}. In light of this, a conservative estimation of the persistence length of an agarose bundle is comparable with that of DNA, \emph{i.e.} {\small$l_p\simeq 50$} nm (this assumes weak attraction between the fibers, see SI). In this case, whose analysis is detailed in the SI, we do not observe significant deviations from the results presented in the main text. It is also worth noticing that this perfectly rigid environment closely resembles artificial gels made of solid nano-wires~\cite{Rahong2014} which possess a much higher Young modulus and have been found to be optimal media for gel electrophoresis experiments.
 

The external field is modelled as a force {\small$\bm{f}$} acting on each bead forming the polymers. Assuming that in physiological conditions half of the charges from the phosphate groups are screened by counter-ions~\cite{Maffeo2010}, one can think that each bead ({\small$\sigma = 2.5$ $nm$ $\simeq 8$ $bp$}) contains a total charge of {\small$q_b = 16q_e/2$}, where {\small$q_e$} is the electron charge. Within this assumption we can map the external force applied onto each bead to an effective electric field {\small$\bm{E} = \bm{f}/q_b$}. Although this mapping is a crude approximation of the Coulomb interaction between the charged DNA, the ions in solution and the applied electric field, we find that we can recover a ``weak field'' behaviour of the knotted samples, \emph{i.e.} linear increase of the speed as a function of their ACN, up to {\small$\sim 50$ $V/cm$}, which is roughly comparable with the field intensity used in experiments. In this work we used field intensities in the range from {\small$E=1.25$ $V/cm$} to {\small$E=625$ $V/cm$}.

The average crossing numbers used in this work have been obtained from  Ref.~\cite{Kusner1994}, where the authors computed the ACN corresponding to M\"obius energy minimising knotted configurations. The thermally averaged ACN of the samples used in this work has been computed from equilibrated configurations and has been found to be in a one-to-one correspondence to the values in Ref.~\cite{Kusner1994} (not shown), confirming the linear relationship between the ACN of ideal and thermally agitated configurations~\cite{Katritch1996a}.

The hydrodynamics is here considered only implicitly, as it is customary for Brownian dynamics simulations. This means that the polymers do not feel one another via hydrodynamical interactions but are subject to thermal fluctuations due to a surrounding bath at fixed temperature {\small$T$} (see SI for more details).

The simulation timescale is given in terms of the Brownian time, which corresponds to the time taken by a bead of size {\small$\sigma$} to diffuse its own size, \emph{i.e.}  {\small$\tau_{Br}=\sigma^2/D_{\sigma}$}, where {\small$D_{\sigma} = k_BT/\xi = k_BT \left(3\pi \eta_{sol} \sigma\right)^{-1}$} is the diffusion coefficient of one bead and {\small$\eta_{sol}=10$ $cP$} the solution (water) viscosity. From this we obtain {\small$\tau_{Br}= 3 \pi \eta_{sol} \sigma^3/k_BT \simeq 40 ns$}.

\end{materials}

\begin{acknowledgments}
DMi acknowledges the support from the Complexity Science Doctoral Training Centre at the University of Warwick with funding provided by the EPSRC (EP/E501311). The computing facilities were provided by the Centre for Scientific Computing of the University of Warwick with support from the Science Research Investment Fund. DMa thanks EPSRC grant EP/I034661/1 for support. EO acknowledges support from the Italian Ministry of Education grant PRIN No.~2010HXAW77.
\end{acknowledgments}

\appendix[Supplementary Information]
\setcounter{figure}{0}
\makeatletter 
\renewcommand{\thefigure}{S\@arabic\c@figure}
\makeatother

\subsection{Model and computational Details}
 \label{app:compdet}

Nicked DNA  knotted loops are modelled as coarse grained bead-and-spring circular chains with a knot manually tied in. Let $\bm{r}_i$ and $\bm{d_{i,j}} \equiv \bm{r}_j - \bm{r}_{i}$  be respectively the position of the center of the $i$-th bead and the vector of length $d_{i,j}$ between beads $i$ and $j$. 

The connectivity of the chain is treated within the finitely extensible non-linear elastic model~\cite{Kremer1990} with potential energy, 
\begin{equation}
U_{FENE}(i,i+1) = -\dfrac{k}{2} R_0^2 \ln \left[ 1 - \left( \dfrac{d_{i,i+1}}{R_0}\right)^2\right]  \notag
\end{equation}
for  $d_{i,i+1} < R_0$ and $U_{FENE}(i,i+1) = \infty$, otherwise; here we chose $R_0 = 1.6$ $\sigma$ and $k=30$ $\epsilon/\sigma^2$ and the thermal energy $k_BT$ is set to $\epsilon$. 
The bending rigidity of the chain is captured with a standard Kratky-Porod potential,
\begin{equation}
U_b(i,i+1,i+2) = \dfrac{k_BT l_K}{2\sigma}\left[ 1 - \dfrac{\bm{d}_{i,i+1} \cdot \bm{d}_{i+1,i+2}}{d_{i,i+1}d_{i+1,i+2}} \right],\notag
\end{equation}
where $l_K = 40 \sigma \simeq 100 nm $ is the known Kuhn length of unconstrained DNA. The steric interaction between beads belonging to the polymers or the gel is taken into account by a truncated and shifted Lennard-Jones potential  
\begin{equation}
U_{LJ}(i,j) = 4 \epsilon \left[ \left(\dfrac{\sigma_c}{d_{i,j}}\right)^{12} - \left(\dfrac{\sigma_c}{d_{i,j}}\right)^6 + 1/4\right] \theta(2^{1/6}\sigma_c - d_{i,j}) \notag.
\end{equation} 
where $\theta(x)$ is the Heaviside function and $\sigma_c$ can be, depending on the case, $\sigma_c = \sigma$ (diameter of the chain beads) and $\sigma_c = (\sigma_g + \sigma)/2 = 5.5$ $\sigma$ where $\sigma_g$ is the diameter of the beads forming the gel. Notice that steric interactions between beads of the gel are excluded from the computation since we consider the gel as a static mesh. This mimics artificial gels made of solid nano-wires, which have Young modulus $10^5$ times larger than that of agarose~\cite{Rahong2014}.

Denoting by $U$ the total potential energy, the dynamic of the beads forming the rings is described by the following Langevin equation:
\begin{equation}
m \ddot{\bm{r}}_i = - \xi \dot{\bm{r}}_i - {\nabla U} + \bm{\eta}
\label{langevin}
\end{equation}
where $\xi$ is the friction coefficient and $\bm{\eta}$ is the stochastic delta-correlated noise. The variance of each Cartesian component of the noise, $\sigma_{\eta}^2$ satisfies the usual fluctuation dissipation relationship $\sigma_{\eta}^2 = 2 \xi k_B T$.

As customary~\cite{Kremer1990}, we set $m/\xi = \tau_{LJ}$, with $\tau_{LJ} = \sigma \sqrt{m/\epsilon}= \sigma \sqrt{m/k_B T}$ and the Brownian time $\tau_{Br}=\sigma/D_b$ with $D_b = k_BT/\xi$ is the characteristic simulation time step. From the Stokes friction coefficient of spherical beads of diameters $\sigma$ we have: $\xi = 3 \pi \eta_{sol} \sigma$ where $\eta_{sol}$ is the solution viscosity. By using the nominal water viscosity, $\eta_{sol}=1$ $cP$ and setting $T=300$ K and $\sigma=2.5$ $nm$, one has $\tau_{LJ} = \tau_{Br} = {3 \pi \eta_{sol} \sigma^3/\epsilon} = 37$ $ns$. Note that, being the gel structure static for all times, no evolution equations for the beads belonging to the mesh are considered. The numerical integration of Eq.~\eqref{langevin} is performed by using a standard velocity-Verlet algorithm with time step $\Delta t = 0.01 \tau_{Br}\sim 0.4$ $ns$ and is implemented in the LAMMPS engine.

The whole system is contained within a box of linear dimension $L=320\sigma$ with periodic boundary conditions in all three directions. The irregularities of a physical gel are created by starting from a regular cubic mesh and then by halving some of the edges randomly, with probability $p=0.4$. The lattice spacing is set to $l=80 \sigma= 200$ $nm$, which is typical for an agarose gel at 5\%~\cite{Pernodet1997}, and the edges of the mesh are discretised with beads of size $\sigma_g = 10 \sigma \simeq 25$ $nm$, again compatible with the size of the agarose bundles~\cite{Pernodet1997,Guenet2006}.

\begin{figure*}[t]
\centering
\includegraphics[width=1.0\textwidth]{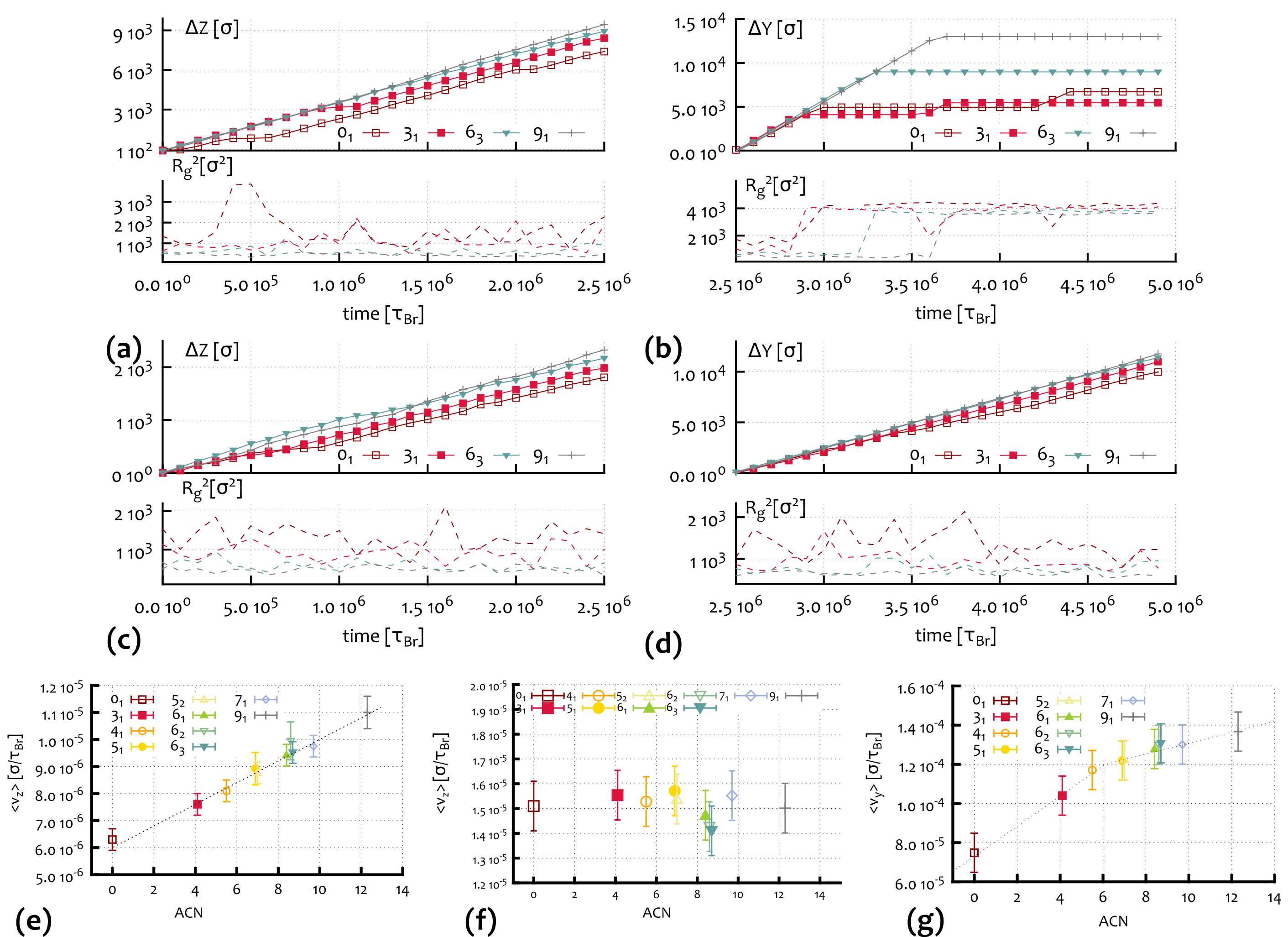}
\caption{Panels \textbf{(a)} and \textbf{(b)}: trajectory and squared radius of gyration vs time for some knotted DNA loops subject to  weak ($50$ $V/cm$) \textbf{(a)}  and moderate ($150$ $V/cm$) \textbf{(b)} electric fields and moving through an irregular gel, \emph{i.e.} with dangling ends. Panels \textbf{(c)} and \textbf{(d)} refer to the regular gel case with same lattice spacing. Note that for high fields and in gel with dangling ends the knotted loops are trapped (zero velocity) for long time in states with large radius of gyration, suggesting an entangled or ``impalement'' state, while in a regular gel the difference between the trajectories at weak and moderate fields is very little.
\textbf{(e)} The velocity of knotted  polymers in regular gels, \emph{i.e.} without dangling ends, displays a monotonic behaviour as a function of the average crossing number. For dense enough gels ($l = 200$ $nm$), the contribution coming from the interaction with the structure is observed to be sufficient to 
recover the linear electrophoretic separation as a function of the average crossing number at weak fields ($E = 1.25$ $V/cm$). This is instead not observed in the case of sparser gels ($l=500$ $nm$), where we recover no electrophoretic separation \textbf{(f)}. We observed a deviation from the linear trend only at stronger fields and denser gels ($E \gtrsim 12.5$ $V/cm$, $l=200$ $nm$), although we never observed a field-inversion or non-monotonic behaviour \textbf{(g)}.}
\label{fig:trj&noDanglingEnds}
\end{figure*}

By considering the beads as cylinders with height and diameter equal to $\sigma$, the system volume fraction results $\phi =  \pi N M \sigma^3/4(L^3 - V_{gel}) \simeq 1.3 \cdot 10^{-4}$, much smaller than the value at which the chains start to overlap ($\phi^{*} \simeq 2 \cdot 10^{-3}$). In other words, the systems are in the dilute limit and interchain interactions occur rarely in the simulations. 

The starting point of the various simulations is a configuration in which simple knotted configurations are placed outside the box. We then slowly pull the rings inside the gel structure, avoiding any impalement. Finally, we place the boundaries of the simulation box so as to match the boundaries of the gel structure and impose periodicity along all the three coordinate directions. To allow the equilibration of the system from this initial state, we first run $1 \cdot 10^7$ $\tau_{Br}$ time steps, which are disregarded before the external force is switched on. The mean and standard error of various properties, as the mean displacement along $\hat{z}$, are calculated by averaging over the rings in the system and over different runs starting from different (equilibrated) initial conditions.

The magnitude of the force $f$ acting on the beads is expressed as a multiple of the system units $\epsilon/\sigma \simeq 1.6 pN$. 
The force acting on the beads can therefore be expressed as the resultant of an electric field $E = 1.6/q_b$ $pN \simeq 12.5$ $kV/cm$. In this work we have used forces $f$ in the range between $10^{-4}$ $\epsilon/\sigma$ and $5$ $10^{-2}$ $\epsilon/\sigma$, which can then be mapped to electric fields roughly between $1.25$ $V/cm$ and $625$ $V/cm$, compatible with the values used in standard DNA gel electrophoresis~\cite{Mickel1977,Levene1987,Viovy2000,Cebrian2014}.

We compute the mobility of knotted polymers by measuring the centre of mass displacement along the direction of the applied field. In Fig.~\ref{fig:trj&noDanglingEnds} we show single example trajectories of the knots and their relative change of radius of gyration during \emph{in silico} gel electrophoresis experiments.
In Fig.~\ref{fig:trj&noDanglingEnds}, we also compare trajectories and speed of knots travelling in regular and irregular gels. As one can notice, while in irregular gels (Fig.~\ref{fig:trj&noDanglingEnds}(a)-(b)) the motion of the knots under weak ($50$ $V/cm$) and moderate ($150$ $V/cm$) fields displays a rather different behaviour, in regular gels (Fig.~\ref{fig:trj&noDanglingEnds}(c)-(d)) knots that are driven by the same external fields display similar trajectories. In particular, we do not observe a field inversion at any field or gel density (Fig.~\ref{fig:trj&noDanglingEnds}(e)-(g)). 

\subsection{The role of Hydrodynamics}
The Brownian dynamics scheme allows us to introduce an implicit solvent, \emph{i.e.} there is no explicit hydrodynamics taken into account during the simulations. We made this choice assuming that in the case of dense gels, the contribution coming from hydrodynamics would be negligible compared to the interactions of the polymers with the environment. This is confirmed in Fig.~\ref{fig:trj&noDanglingEnds} where we show the linearly increasing speed of knots travelling through a regular gel at weak and strong fields. In both cases, even without the presence of hydrodynamics or rescaling, we recover the expected linear relationship between speed and ACN. In Ref.~\cite{Weber2006a} the authors rescaled the final velocities of the knots by using the Kirkwood-Risenman formula, thereby accounting for hydrodynamics. In this work we observed that such rescaling does not affect the final functional form of the results, consequently we did not introduce such rescaling. We argue that the contribution from hydrodynamics becomes relevant when sparser gels are considered. In this case we observed (data not shown) that the rescaling operated in Ref.~\cite{Weber2006a} is necessary to recover the experimental observations as no electrophoretic separation is observed (see Fig.~\ref{fig:trj&noDanglingEnds}(f)).

\begin{figure*}[t]
\centering
\includegraphics[width=0.78\textwidth]{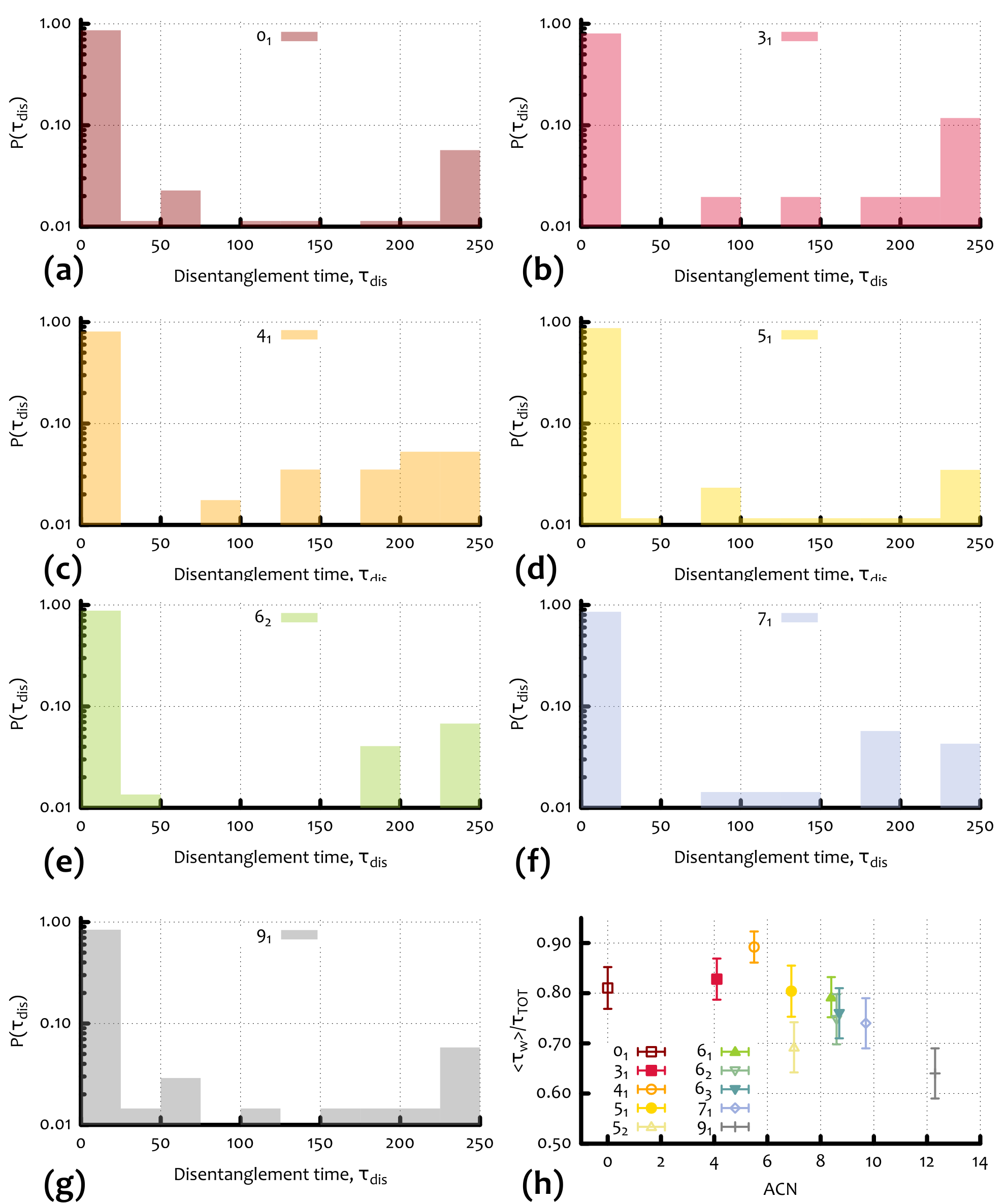}
\caption{\textbf{(a)-(g)} Waiting time distribution $P(\tau_{\rm dis})$ for different knot types and in units of simulation Brownian time $\tau_{Br}$. One can notice that the distribution shows to peaks at short and large times which correspond to ``transverse'' or ``head on'' interactions, respectively. \textbf{(h)} Total amount of time spent in an entangled state as a function of the ACN. This quantity reflects the fact that slower knots spend more time in entangled states (compare with Fig.2 in the main text).}
\label{fig:Panel_Wtime}
\end{figure*}  

\subsection{Waiting Times}

In order to quantify the time taken by the rings to disentangle from the impalements, we analyse the single trajectories (reported for example in Fig.~\ref{fig:trj&noDanglingEnds}) and compute the amount of time each knot is moving slower than it would if it were free, \emph{i.e.} $v_{free}=f/\zeta$. From this we can quantify the total amount of time each knot is stuck in an entangled state. This is found to be a non-monotonous function of the ACN (see Fig.~\ref{fig:Panel_Wtime}(h)) and is in line with the non-monotonic average speed of the knots and the relative separation in gel electrophoresis experiments as a function of their ACN.

Each time a knot is stalled, it takes a certain amount of time to re-establish its motion. We report the distribution of these ``waiting times'' ($\tau_{\rm dis}$) in Fig.~\ref{fig:Panel_Wtime}(a)-(g) for several knots. As one can notice, the distribution is bimodal, \emph{i.e.} shows two distinct peaks at short and long times. We interpret the peak at short time-scales as transversal impalements or crawling around the gel structure, while the peak at long times as the fact that sometimes knots interact via an ``head-on'' collision with the gel open strands. This immobilises the knots for long time, as the only way to re-establish the motion is by moving against the field for a length of at least half lattice spacing.

\subsection{Biased Continuous-Time Random Walk Model}
\label{app:Timescales}

Here we detail the biased continuous time random walk (CTRW) model, whose results are reported in the main text.

We start by noticing that the simulations suggest the existence of two important time-scales in the dynamics: 
$\tau_f$ and $\tau_{\rm dis}$, respectively regulating the frequency of entangling events and the time required to 
disentangle from the impalements with the dangling ends.  
Firstly, we address the situation in which $\tau_{\rm dis} \ll 1$. In this case the knots disentangle quickly, 
and we can think of the whole process as random walks biased in the opposite (the DNA is negatively charge) direction of the field. 
Due to topological interactions with the gel, the biased random walkers eventually stop with a rate $\lambda_e \simeq \tau_f^{-1}$. 
We expect that this rate should be described by the probability $P_e(r/l)$ of a circle of radius $r$ to overlap 
onto an edge of 2D square mesh with lattice spacing $l$, \emph{i.e.}
\begin{equation}
\lambda_e(\mathcal{K}) \simeq \dfrac{P_e(x(\mathcal{K}))}{P_e(x(0_1))}\lambda_e(0_1),
\end{equation}
where $x(\mathcal{K})=R_g(\mathcal{K})/l$.
Note that this rate is normalised with respect to the empirical value observed in the Brownian dynamics simulations for the unknot ($0_1$) at weak fields.

\begin{figure*}[t]
\centering
\includegraphics[width=0.45\textwidth]{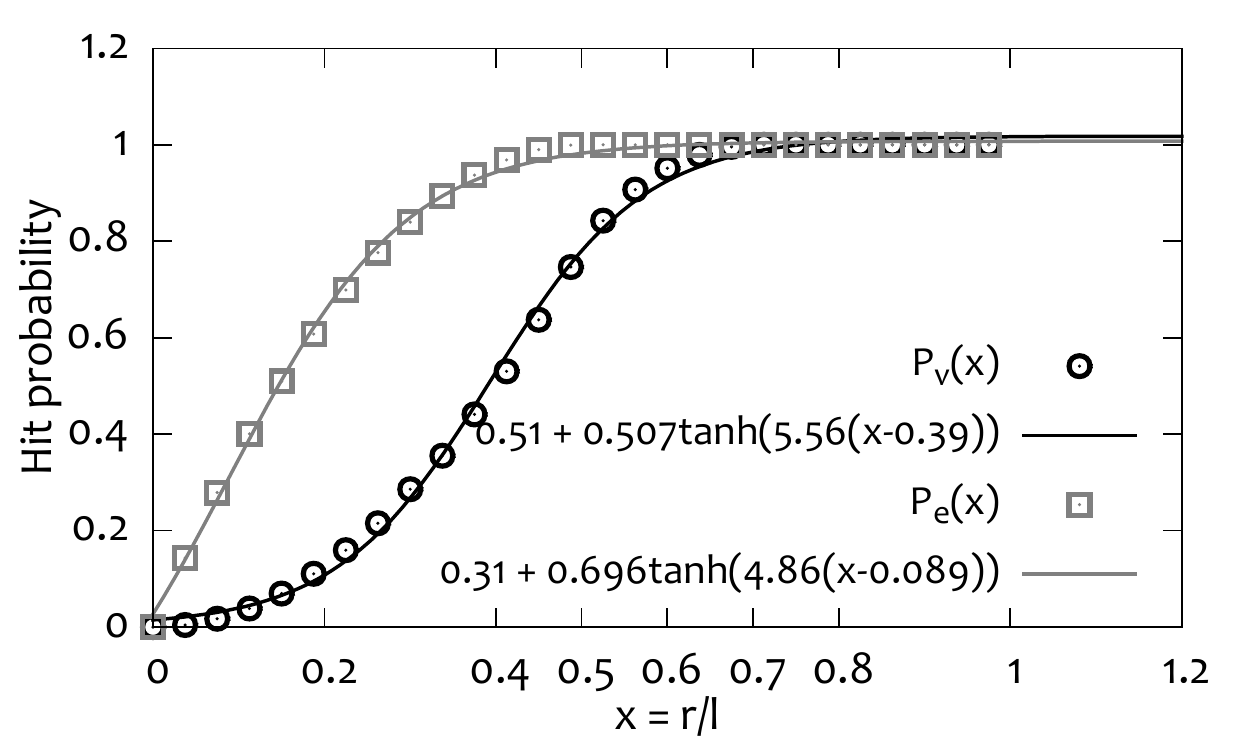}
\caption{Probability for a circle of radius $r$ drawn randomly on a square lattice with lattice spacing $l$ to encircle a lattice vertex ($P_v$ -- black circles) and to overlap over a lattice edge ($P_e$ -- gray squares). The two quantities reach 1 for $r=l/2$ ($P_e$) and for $r = l/\sqrt{2}\simeq 0.7$ ($P_v$). The functions and parameters used to fit the data are also shown.}
\label{fig:hitprob}
\end{figure*}

We can obtain the value of $P_e(x)$ as a function of the ratio $R_g(\mathcal{K})/l$, with $l$ the gel lattice spacing from Fig.~\ref{fig:hitprob}, where we simulated the deposition of a circle of radius $r$ onto a 2D mesh with lattice spacing $l$ and calculated the probability of overlapping over a strand ($P_e(x)$) or to encircle a vertex ($P_v(x)$) of the grid. 

\begin{figure*}[t]
\centering
\includegraphics[width=0.95\textwidth]{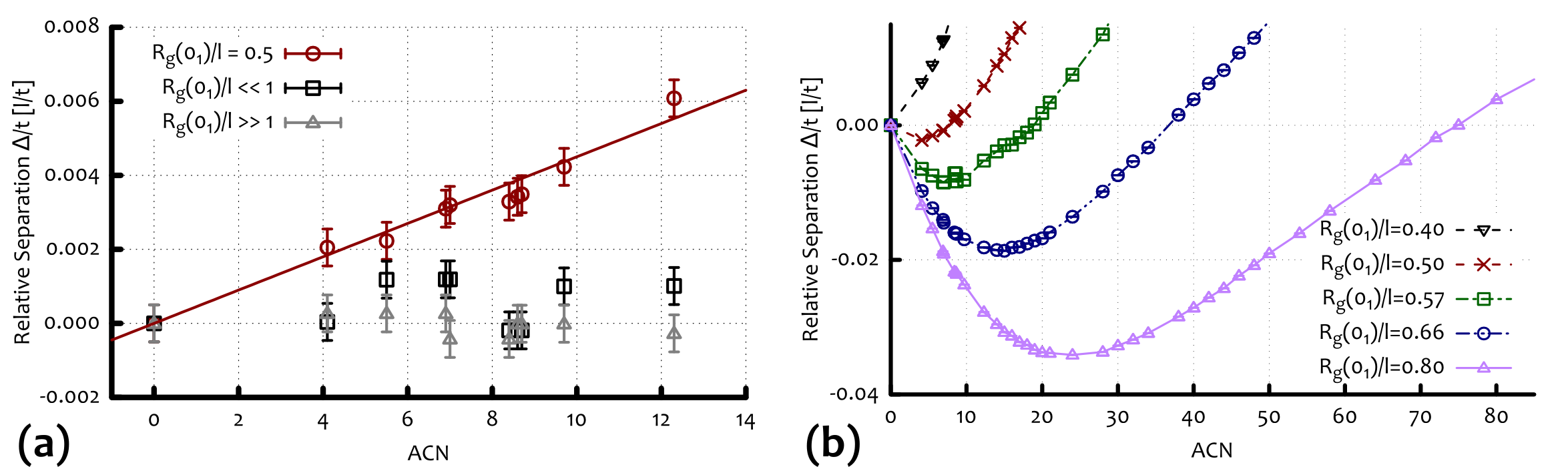}
\caption{Average relative separation over time (in units of lattice spacing over time) of the knots as a function of the ACN obtained from the simple CTRW model described in the text. \textbf{(a)} For weak fields (for which the disentanglement time is quick $\tau_{\rm dis}\ll 1$) the linear relationship is re-established for $R_g \sim l$, while in the two limiting cases $R_g \gg l$ and $R_g \ll l$ we do not observe a clear separation, as expected. \textbf{(b)} The non-monotonic behaviour of the separation is re-established when a topology-dependent disentanglement time is introduced in the model. The typical electrophoretic arc crucially depends on the ratio $R_g(ACN)/l$. (See text for details).}
\label{fig:RWmodel}
\end{figure*}

The knots moving in the gel are therefore viewed as random walkers on a lattice. They can jump one site to their right (broadly this corresponds to one unit cell of the gel), or are required to stop for one time-step (because $\tau_{\rm dis}\ll 1$) with rate $\lambda_e$. This is equivalent to a Poisson process whose probability of getting entangled after $t$ time-steps is given by:
\begin{equation}
P_{ent}(x(ACN),t) = 
1 - [1 - \lambda_e(R_g(\mathcal{K}))]^t \simeq 1- \exp{[-\lambda_{e}(\mathcal{K}) t]},
\end{equation}
where we stress the dependence of $\lambda_e$ on the knot type via its size $R_g$. The average time between interactions is given by $\langle \tau_f\rangle \simeq \lambda_e^{-1}$ and the average number of stops on a time $t$ is $\langle n_e \rangle = \lambda_e t$.
The relative separation between two knots can be expressed as
\begin{align}
\Delta x(\mathcal{K}_1,\mathcal{K}_2) &= v_{free} (t_{free}(\mathcal{K}_1) - t_{free}(\mathcal{K}_2)) \notag \\
&= v_{free} t (\lambda_e(\mathcal{K}_2) - \lambda_e(\mathcal{K}_1)) 
\end{align}
where $v_{free}$ is the velocity of the knots when no interactions occur (when hydrodynamics is negligible this is the same for all knots), $t$ is the observation time and $t_{free} = t - \lambda_e t$. From this is clear that the separation between knots $\mathcal{K}_2$ and $\mathcal{K}_1$ is proportional to the difference of their entanglement rates, as expected.
Given that our model relies on the hitting probability between a knot and its environment, it is clear that if the knot size is either much larger or much smaller than the gel lattice spacing, the electrophoretic separation is not as good as in the case the two are comparable. Within our simple model we recover good linear relationship within the range $R_g/l \in [10^{- 4}, 10]$ meaning that short samples (or large gel pores compared to the gyration radius of the knots) work better for the electrophoretic separation at weak fields rather than small pores (or very large $R_g$)~\cite{Dorfman2010} (see Fig.~\ref{fig:RWmodel}(a)). This is compatible with the fact that is often impractical to perform gel electrophoresis with very large samples. In fact, for other cases outside this range, we  observe essentially no spatial separation of the knots (see Fig.~\ref{fig:RWmodel}(a)).




We now introduce a topology dependent disentanglement time: every time that the Random Walker undergoes a ``stop'', we require that it has to wait an amount of time that increases proportionally to its average entanglement number (see $\langle \pi \rangle$ in Fig.~3 of main text). 
We pick the disentanglement time from an exponential distribution with characteristic time given by the empirical one observed in the MD simulations (see Fig.~4 in the main text), \emph{i.e.} 
\begin{equation}
t_{w} = -\ln{r}/\lambda_{\rm dis} 
\end{equation}
where $r$ is a random number and $\lambda_{\rm dis}^{-1} = \tau_{\rm dis}$. As observed in the main text, the disentanglement time $\tau_{dis}$ is expected to be a function of $\langle \pi\rangle$, \emph{i.e.} to increase linearly with the ACN. We therefore use:
\begin{equation}
\tau_{dis}(ACN) = A\langle \pi \rangle +\tau_{\rm dis}(0_1) 
\end{equation}
where $A=5$ $\tau_{Br}$ is an empirical parameter fitted from the simulations and $\tau_{dis}(0_1)=30 \tau_{Br}$ is the empirical average disentanglement time of the unknot. In addition, long waiting times which are typical of head-on impalements are taken into account by adding a small probability ($q \simeq 0.05$, compatible with the bimodal distribution in Fig.~4 of the main text) and conditioned to the fact that an entanglement event happened. For this we set $\tau_{long} = A\langle \pi \rangle + 250$. So the average disentanglement time from a head-on impalement is still knot-dependent but is much longer than any characteristic disentanglement time for other types of entanglements. 
 
Within this model we recover the electrophoretic arc in Fig.~\ref{fig:RWmodel}(b). In addition, the shape of the arc can be finely tuned to obtain different depths of the arc and different slowest knots.  This is mainly controlled via the ratio $R_g/l$ (we here assume that the external field is neither too weak or too strong). 
In Fig.~5 of the main text we show that within this model we can accurately recover the electrophoretic speed of the knots observed in the MD simulations by shifting the curve obtained with the CTRW model setting $R_g(0_1)/l=0.5$ (as in the MD simulation) to the observed speed of the unknot and by rescaling it by the free speed $v_{free}$ at which the knots travel between two planes of the gel, \emph{i.e.} two lattice sites in the CTRW model.  
It is worth noting also that this model allows us to predict the mobility of knots which were not probed by Brownian dynamics, by making use of the empirical functional forms of $\langle \pi \rangle$ and $R_g$ derived from these simulations. It is also worth stressing that the only real free parameter of this model is the lattice spacing $l$, while the value of $R_g(0_1)$, $\tau_{\rm dis}(0_1)$ and $\lambda_e(0_1)$ are input parameters needed to calibrate the model to a target case.

\subsection{Gel with flexible Dangling Ends}
While the assumption that the gel is perfectly rigid is only an approximation for agarose gels, several experiments reported values for the persistence length of single agarose fibers in the range of $2-10$ nm; in particular Ref~\cite{Guenet2006} reports $l_{\rm agarose} = 9$ nm (the same work also presents evidence of the presence of dangling ends in the gel). Furthermore, each agarose bundle (or ``fibril'') is formed by a number $n_f$ of fibers, with $n_f$ ranging between 10 and 20~\cite{Pernodet1997,Guenet2006}, and has a diameter $\sigma_f \simeq 10 - 20$ nm~\cite{Sugiyama1994}. The persistence length of a full fibril might depend linearly or quadratically on $n_f$ (depending on the structure of the bundle and strength of interfiber interactions, see Ref.~\cite{Mogilner2005}). Therefore, a conservative estimate of the persistence length of an agarose bundle is attained if a linear scaling is assumed (corresponding to weak attractive interactions between the fibers, see Ref.~\cite{Mogilner2005}), leading to 
$l_{\rm agarose} \sim l_{\rm DNA} \simeq 50$ nm.

%

\begin{figure*}[t]
\centering
\includegraphics[width=0.9\textwidth]{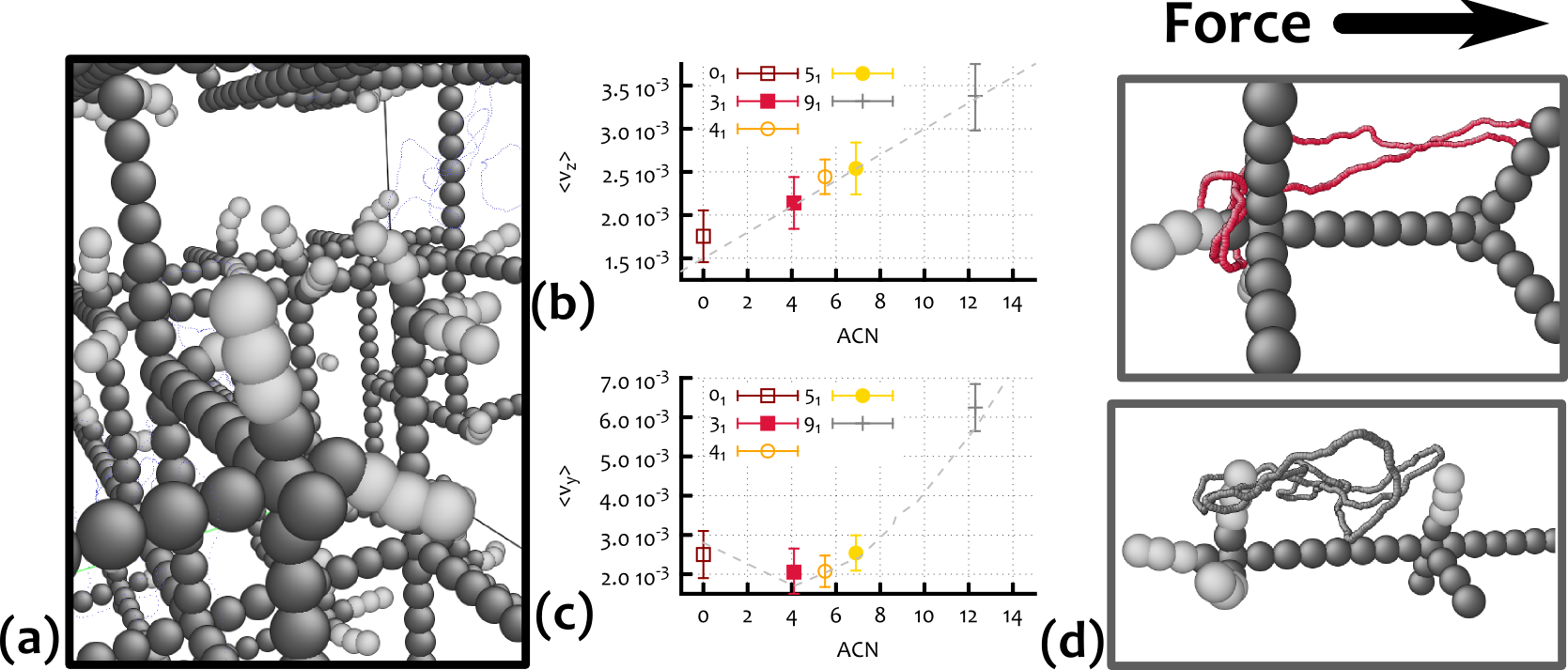}
\caption{\textbf{(a)} Picture of the gel with flexible dangling ends (light grey) anchored to the rigid structure (dark grey). \textbf{(b)-(c)} Average velocity of some knots under weak ($50$ $V/cm$) and moderate ($150$ $V/cm$) field, respectively.  \textbf{(d)} Snapshots (not to scale) of entanglement events of $3_1$ (top) and $9_1$ (bottom) knots corresponding to stalling points in their trajectories and large values of $R_g$.}
\label{fig:SI_flxDE}
\end{figure*}

In order to probe the effect of fibers' flexibility on the entanglement/disentanglement 
process experienced  by the knots travelling through the gel, we therefore modify the gel structure to accommodate flexible dangling ends (see Fig.~\ref{fig:SI_flxDE}(a)) described as semi-flexible polymers with  persistence length equal to that of dsDNA, and anchored to cross-links that are still modelled as rigid and static. This choice corresponds to flexible agarose bundles clamped at right angles at crosslinks~\footnote{The simulations in Fig.~\ref{fig:SI_flxDE} consider flexible bundles with fixed angles at crosslink, i.e. the clamping at the crosslinks is infinitely strong. For the case of the unknot, we checked that replacing this right angle constraint with a soft angular potential keeps the dynamics qualitatively unchanged (we still observe entanglements or impalements).}, and avoids the un-physical scenario in which the moving molecule would drag the gel along. 
As one can see from  Fig.~\ref{fig:SI_flxDE}(b)-(c), the qualitative behaviour of knots dynamics does not change with respect to the fully rigid gel. In particular, as soon as the external electric field assumes moderate values, the non-monotonic behaviour of the average velocity is recovered. 
In Fig.~\ref{fig:SI_flxDE} we also show some snapshots of knots with relative large values 
of (instantaneous) $R_g$  that become entangled with the flexible dangling ends: correspondingly their trajectories stall.
 
From the above analysis we conclude that, while adding flexibility to the gel network  might 
change the numerical values of, for instance, the disentanglement time of the knots, the qualitative non-monotonic response as a function of their ACN, when driven by a moderate field, is not substantially changed. \\
\\
\bibliographystyle{pnas}
\bibliography{EP_knots,MyRefs,Books}

\end{article}
\end{document}